\documentclass[english,journal]{IEEEtran}
\usepackage[T1]{fontenc}
\usepackage{color}
\usepackage{babel}
\usepackage{float}
\usepackage{bm}
\usepackage{amsmath}
\usepackage{amssymb}
\usepackage{graphicx}
\usepackage[unicode=true,
 bookmarks=true,bookmarksnumbered=true,bookmarksopen=true,bookmarksopenlevel=1,
 breaklinks=false,pdfborder={0 0 0},pdfborderstyle={},backref=false,colorlinks=false]
 {hyperref}
\hypersetup{pdftitle={Your Title},
 pdfauthor={Your Name},
 pdfpagelayout=OneColumn, pdfnewwindow=true, pdfstartview=XYZ, plainpages=false}

\makeatletter

\floatstyle{ruled}
\newfloat{algorithm}{tbp}{loa}
\providecommand{\algorithmname}{Algorithm}
\floatname{algorithm}{\protect\algorithmname}

\usepackage{cite}
\ifCLASSOPTIONcompsoc
\usepackage[caption=false,font=normalsize,labelfont=sf,textfont=sf]{subfig}
\else
\usepackage[caption=false,font=footnotesize]{subfig}
\fi
\usepackage{psfrag,cite}
\usepackage{algorithmic}
\ifCLASSOPTIONcompsoc
\usepackage[caption=false,font=normalsize,labelfont=sf,textfont=sf]{subfig}
\else
\usepackage[caption=false,font=footnotesize]{subfig}
\fi

\usepackage{tikz} 
\usepackage{pgfplots} 
\pgfplotsset{compat=newest} 
\pgfplotsset{plot coordinates/math parser=false} 
\newlength\figureheight 
\newlength\figurewidth 
\usepackage{tikz}
\usepackage{tikzscale}

\makeatother

\begin{document}

\title{Transmitter Beam Selection in Millimeter-wave MIMO with In-Band Position-Aiding}

\author{Gabriel~E.~Garcia, Gonzalo Seco-Granados,~\IEEEmembership{Member,~IEEE},
Eleftherios Karipidis, \\and Henk Wymeersch,~\IEEEmembership{Member,~IEEE}\thanks{G.~E.~Garcia, and H.~Wymeersch are with the Department of Electrical
Engineering, Chalmers University of Technology, Gothenburg, Sweden,
e-mails: \{ggarcia,henkw\}@chalmers.se. G.~Seco-Granados is with
the Department of Telecommunications and Systems Engineering, Universitat
Autònoma de Barcelona, Barcelona, Spain, e-mail: gonzalo.seco@uab.cat.
E.~Karipidis is with Ericsson Research, Stockholm, Sweden, e-mail:
eleftherios.karipidis@ericsson.com. This research was supported, in
part, by the European Research Council, under Grant No. 258418 (COOPNET),
the EU project HIGHTS (High precision positioning for cooperative
ITS applications) MG-3.5a-2014-636537, and the R\&D Project of Spanish
Ministry of Economy and Competitiveness under Grant TEC2014-53656-R.}}
\maketitle
\begin{abstract}
Emerging wireless communication systems will be characterized by a
tight coupling between communication and positioning. This is particularly
apparent in millimeter-wave (mm-wave) communications, where devices
use a large number of antennas and the propagation is well described
by geometric channel models. For mm-wave communications, initial access,
consisting in the beam selection and alignment of two devices, is
challenging and time-consuming in the absence of location information.
Conversely, accurate positioning relies on high-quality communication
links with proper beam alignment. This paper studies this interaction
and proposes a new position-aided beam selection protocol, which considers
the problem of joint communication and positioning in scenarios with
direct line-of-sight and scattering. Simulation results show significant
reductions in latency with respect to a standard protocol.
\end{abstract}

\section{Introduction\label{sec:Introduction}}

\IEEEPARstart{M}{illimeter-wave} (mm-wave) communications have
recently gained attention for the development of high-speed wireless
networks. Mm-wave systems operate at frequencies between 30 to 300
GHz with large available bandwidths. Combined with multiple-input-multiple-output
(MIMO), using a large number of antennas, mm-wave can provide high
data rates to users through dense spatial multiplexing \cite{Zhouyue,Rappaport,Swindlehurst2014,Sun2014,Bai2014}.
Hence, mm-wave MIMO is considered a key enabler for emerging communication
systems, e.g., 5G or IEEE WiGig \cite{Hansen2011}, to deliver throughputs
on the order of multi-Gbps for a range of applications from wearables
\cite{Venugopal2016} to automotive \cite{Choi2016}. However, mm-wave
communications face a number of challenges, in particular severe path-loss
at these high frequencies. As a solution, system designers improve
the link budget through highly directional links involving sophisticated
beamforming (BF) at the transmitter and/or receiver \cite{Wang,Hur,Tsang},
relying on the knowledge of the mm-wave MIMO propagation channel.

Given the quasi-optical propagation of mm-wave, stochastic geometrical
channel models have become an attractive approach to characterize
the channel with few parameters. These models relate the propagation
to the geometry of the operating environment, thus creating an explicit
interplay between the communication channel and the positions of the
transmitter, receiver, and reflectors \cite{Lopes1995,DenSaya,Li2014,Leus,Steinbauer2001}.
This interplay becomes apparent during the initial access phase, where
two devices, a transmitter and a receiver, here termed D1 and D2,
aim to establish a connection by achieving beam alignment. This consists
in finding a pair of transmit and receive beams to reach a required
signal-to-noise ratio (SNR) for the link. From the communications
perspective, this is achieved by a dedicated protocol that searches
across the angle-of-arrival (AOA) and angle-of-departure (AOD) space.
Both AOA and AOD can be related to the location of D1 and D2, thus
presenting an opportunity to exploit location information. From the
 positioning perspective, obtaining the position information of a
device through exchange of mm-wave signals requires the establishment
of a communication link. Hence, the communication and positioning
problems are coupled, indicating that a joint solution strategy may
yield better performance. 

Conventional beam selection protocols do not consider the positioning
aspect explicitly. For instance, the authors in \cite{Wang,Lee2011,Xia2008,Chen2011,Xiao2016}
designed BF protocols based on discretized iterative beam codebooks,
while in \cite{Tsang} the use of simultaneous beams through beam
coding is introduced. In \cite{Leus,Kokshoorn2017}, the authors developed
a hierarchical multi-resolution codebooks: in \cite{Leus}, codebooks
are based on hybrid analog/digital precoding and proposed low-overhead
channel estimation algorithms, while in \cite{Kokshoorn2017} the
codebook allows for beam overlapping for channel estimation purposes.
In \cite{Barati2016}, the initial access problem is tackled by means
of scanning and signaling procedures, while in \cite{Liu2017} the
authors propose a strategy for transmitting reference signals using
pre-designed codebooks for device discovery, and in \cite{Kim2014},
prioritized beam ordering strategies are presented. These protocols
involve a time-consuming search over different AOA/AOD pairs in order
to determine directions in which to point the beams. On the other
hand, contributions in the area of  positioning generally ignore the
initial access aspect. For instance, the authors in \cite{hu2014esprit,DenSaya,sanchis2002novel,Zhu2016,Peng2016}
present direction-of-arrival and location estimation algorithms, but
do not provide initial access protocols. Similarly, \cite{Shamansoori2015}
exploits mm-wave and MIMO features along with BF to provide sufficient
conditions on the identifiability of the position and orientation
for a device in a line-of-sight (LOS) scenario but no protocols for
the initial access are included. Works that combine positioning with
initial access include \cite{Alexandropoulos2017,Aviles2016,Garcia2016,Va2015}:
\cite{Alexandropoulos2017} proposes a beam alignment method for fixed-position
network nodes in mm-wave backhaul systems aided with position information
obtained using high-sensitivity displacement sensors in each node.
In \cite{Aviles2016}, beam training is presented exploiting a database
linked to the geographical position of the users. In \cite{Garcia2016}
location information is harnessed for fast channel estimation in a
vehicular context. In \cite{Va2015}, beam alignment is proposed with
the use of position information obtained from the on-board train system.
What is common in \cite{Alexandropoulos2017,Aviles2016,Garcia2016,Va2015}
is that position information is obtained out-of-band, not from the
mm-wave signal itself. In the context of beam tracking (i.e., once
the initial access has been solved) in-band information has been harnessed,
in the form of either AOD or/and AOA \cite{Cao2016,Bae2017,Kela2016}:
authors in \cite{Bae2017} propose an estimator for the AOD and channel
information under Gaussian AOD dynamics, but no protocol is presented;
in \cite{Cao2016}, AOA estimation is introduced based on the geometry
of the antenna array and the transmitting beam pattern, not including
position information; in \cite{Kela2016}, state-space models for
the AOD and AOA are inferred aided with channel-aided information
rather than position information.

In this paper, a novel in-band positioning-aided transmitter beam
selection protocol is proposed, with the aim of reducing the set-up
time of the initial access procedure for communication in the presence
of a line-of-sight path and unknown scatterer locations. In order
to gain insight into the fundamental achievable performance, we determine
the evolution of the Fisher information of the D2 position and orientation
as new beams are utilized, feeding back this location information
to D1 in order to adapt the beams. Both discrete and continuous codebooks
are considered, in the presence of LOS communication with scatterers.
The new protocol is evaluated through simulations, considering as
performance metrics the set-up time, signal-to-noise (SNR) ratio,
and the position and orientation error bounds after protocol completion.
We observe that the position-aided protocol is significantly faster
than a conventional protocol based on discretized beam codebooks,
with little or no SNR penalty, and can additionally determine the
position or orientation of D2. In addition, we find that standard
discrete codebooks achieve similar performance to more complex codebooks,
indicating that the proposed protocol can be implemented with standard
mm-wave communication technologies. 

The remainder of the paper is structured as follows. Section \ref{sec:System-Model}
presents the communication model and performance metrics. In Section
\ref{sec:Conventional-Beam-Selection}, the conventional protocol
description, operation and performance are described. Then, in Section
\ref{sec:Joint-PBS} the joint positioning and beam selection protocol,
its operation and performance are introduced. Finally, numerical results
are given in Section \ref{sec:Simulation-Results}, followed by the
conclusions in Section \ref{sec:Conclusions}.

\section{System Model\label{sec:System-Model}}

\subsection{Geometric Model}

\begin{figure}[t]
\begin{centering}
\includegraphics[width=1\columnwidth]{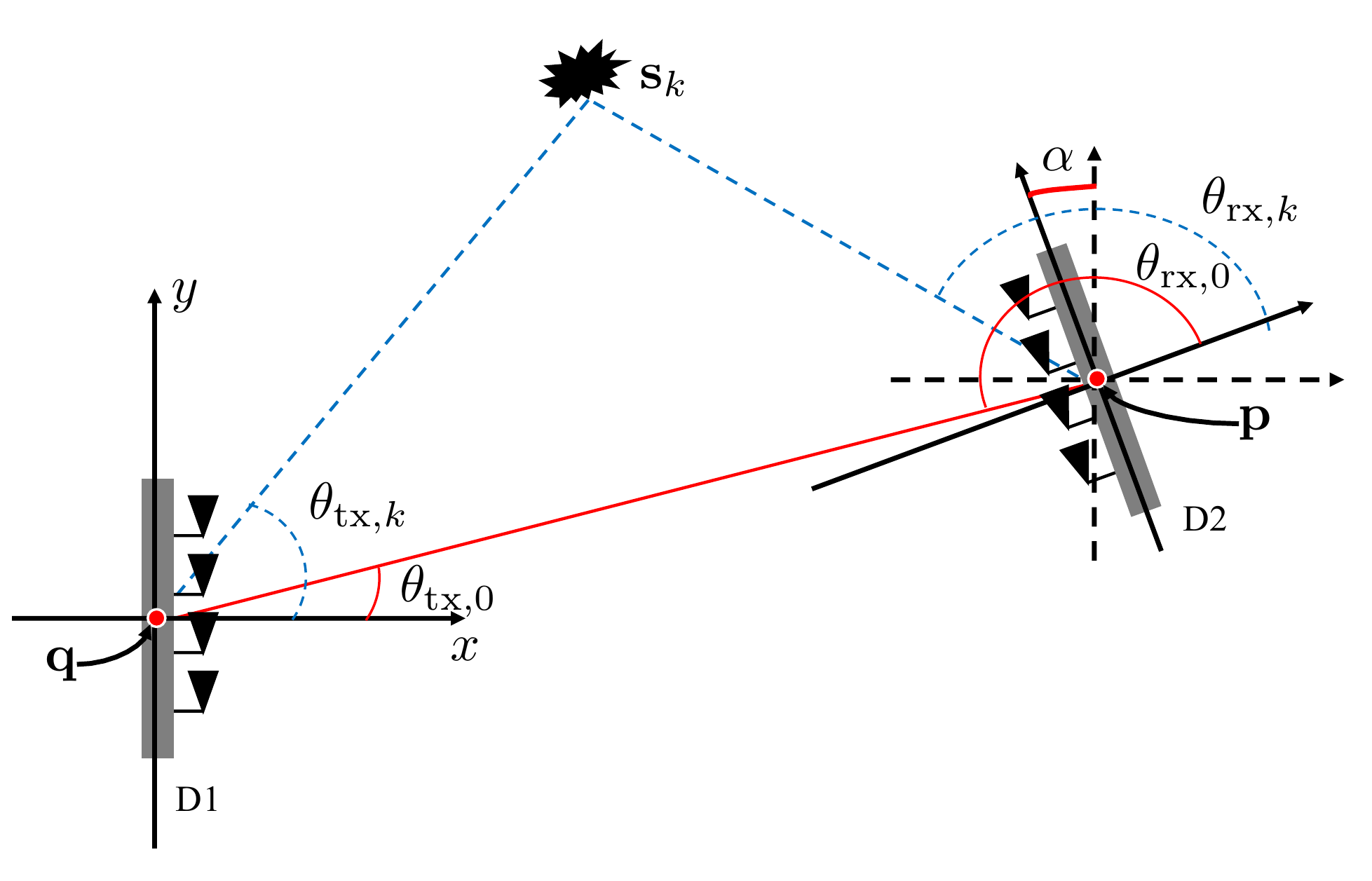}
\par\end{centering}
\caption{Two-dimensional MIMO system model with a D1 with known position and
orientation, and a D2 with unknown position ($\mathbf{p}$) and orientation
($\alpha$). The AODs $\theta_{\mathrm{tx},k}$ and AOAs $\theta_{\mathrm{rx},k}$
are also indicated. \label{fig:system_model}}
\end{figure}

We consider a MIMO mm-wave system consisting of a transmitting device
D1 with $N_{t}$ antennas and beamforming capabilities, and a receiving
device D2 with $N_{r}$ antennas. The 2-dimensional locations\footnote{A 2-dimensional model is assumed for simplicity. However, the proposed
protocols can be extended to 3-dimensional scenarios with 2-dimensional
antenna arrays.} of D1 and D2 are denoted by $\mathbf{p}=[p_{\mathrm{x}},p_{\mathrm{y}}]^{\mathrm{T}}\in\mathbb{R}^{2}$
and $\mathbf{q}=[q_{\mathrm{x}},q_{\mathrm{y}}]^{\mathrm{T}}\in\mathbb{R}^{2}$,
respectively, and let $\alpha\in[0,2\pi)$ be the angle of rotation
of the D2 antenna array with respect to the horizontal axis. These
parameters in turn imply an AOD $\theta_{\mathrm{tx},0}$ and an AOA
$\theta_{\mathrm{rx},0}$, as depicted in Figure \ref{fig:system_model}.
Note that under our definitions, $\cos(\theta_{\mathrm{tx},0})=(p_{\mathrm{x}}-p_{\mathrm{y}})/\left\Vert \mathbf{q}-\mathbf{p}\right\Vert ,$
and $\alpha=\pi+\theta_{\mathrm{tx},0}-\theta_{\mathrm{rx},0}.$ We
also introduce the LOS propagation delay as between D2 and D1 as $\tau_{0}=\left\Vert \mathbf{q}-\mathbf{p}\right\Vert /c$,
where $c$ is the speed of light. We assume that $\mathbf{q}$ is
a known reference point. It is easy to show that the knowledge of
$\bm{\beta}=[\mathbf{p},\alpha]^{\mathrm{T}}$ is equivalent to the
knowledge of $[\tau_{0},\theta_{\mathrm{tx},0},\theta_{\mathrm{rx},0}]$.
The environment can also contain scatterers, here modeled as points,
with locations $\mathbf{s}_{k}$, $k\ge1$, for which we introduce
$\tau_{k}=\left\Vert \mathbf{q}-\mathbf{s}_{k}\right\Vert /c+\left\Vert \mathbf{s}_{k}-\mathbf{p}\right\Vert /c$,
as well as the AOD $\theta_{\mathrm{tx},k}$ and AOA $\theta_{\mathrm{rx},k}$,
as shown in Figure \ref{fig:system_model}. 

Furthermore, we consider that the device D1 transmits signals at a
carrier frequency $f_{c}$ (or equivalently wavelength $\lambda=c/f_{c}$,
where $c$ is the speed of light) and with bandwidth $B$. We employ
a narrowband model\footnote{The narrowband assumption imposes the constraints that (i) there is
no beam squint; and (ii) $\mbox{max}(N_{t},N_{r})d\ll c/B$ where
$d$ denotes the distance between the antenna elements. } where the $N_{r}\times N_{t}$ channel matrix is given by \cite{Sayeed2002,BradySay2}
\begin{equation}
\mathbf{H}(t)=\sum_{k=0}^{K-1}\underbrace{\sqrt{N_{t}N_{r}}\,h_{k}~\mathbf{a}_{\mathrm{rx}}(\theta_{\mathrm{rx},k})\mathbf{a}_{\mathrm{tx}}^{\mathrm{H}}(\theta_{\mathrm{tx},k})}_{=\mathbf{H}_{k}}\delta(t-\tau_{k}),\label{eq:channelmodel}
\end{equation}
in which $h_{k}$ is the complex gain of the $k$-th path, $\mathbf{a}_{\mathrm{tx}}(\theta_{\mathrm{tx},k})\in\mathbb{C}^{N_{t}}$
and $\mathbf{a}_{\mathrm{rx}}(\theta_{\mathrm{rx},k})\in\mathbb{C}^{N_{r}}$
are the normalized antenna steering and response vectors associated
with the $k$-th path. 

Without loss of generality, our focus will be on uniform linear arrays\footnote{The underlying idea of the proposed protocol is applicable to any
array geometry.} (ULA), for which 
\begin{eqnarray}
\left[\mathbf{a}_{\mathrm{tx}}(\theta_{\mathrm{tx}})\right]_{l=0}^{N_{t}-1} & = & \frac{1}{\sqrt{N_{t}}}\exp\bigl(j\frac{2\pi ld}{\lambda}\sin\theta_{\mathrm{tx}}\bigr)\label{eq:steeringvector}\\
\left[\mathbf{a}_{\mathrm{rx}}(\theta_{\mathrm{rx}})\right]_{l=0}^{N_{r}-1} & = & \frac{1}{\sqrt{N_{r}}}\exp\bigl(j\frac{2\pi ld}{\lambda}\sin\theta_{\mathrm{rx}}\bigr),\label{eq:responsevector}
\end{eqnarray}
where $d$ is the antenna spacing. 

\subsection{Training Model\label{subsec:Training-Model}}

\subsubsection{Transmitter With Analog Beamforming}

We assume the use of analog beamforming, implemented with phase shifters
and combined with antenna selection. The transmitter D1 can sequentially
send training sequences (TS) using beams pointed towards in different
directions, leading to a signal model $\mathbf{f}_{m}x(t)$ in which
$x(t)=\sum_{n=1}^{N}a_{n}p(t-n/B),$ where $p(t)$ is a unit-energy
transmit pulse (e.g., root-raised cosine), $N$ denotes the number
of symbols, and $a_{n}$ are known training symbols with $\mathbb{E}\{|a_{n}|^{2}\}=E_{s}$,
and 
\begin{align}
 & \mathbf{f}_{m}=\\
 & \frac{1}{\sqrt{N'_{t}}}\left[\mathbf{0}_{\frac{N_{t}-N'_{t}}{2}}\,e^{j\phi_{0}}\ldots e^{j\phi_{N'_{t}-1}}\,\mathbf{0}_{\frac{N_{t}-N'_{t}}{2}}\right]^{\mathrm{T}},\nonumber 
\end{align}
where $N'_{t}\le N_{t}$ indicates the number of\emph{ active contiguous
antennas} \cite{Celik2006} used to control the beam widths at the
expense of the beam gain. Special cases include 
\begin{equation}
\phi_{i}\in\{0,\pi,\pi/2,-\pi/2\}\label{eq:simpleCodebook}
\end{equation}
and 
\begin{equation}
\phi_{i}=e^{j\frac{2\pi di}{\lambda}\sin\theta_{m}},\label{eq:directionCodebook}
\end{equation}
where $\theta_{m}$ is the direction of the beam, chosen from a given
set $\Theta$. The design parameters of the beam patterns consist
of the maximum gain direction $\theta_{\mathrm{max}}$ and the half-power
beam-width angle, $\theta_{\mathrm{HPBW}}$, which is the angle where
the square magnitude of the radiation pattern decreases by 50\% with
respect to its maximum value, and depends on the type of antenna and
operating frequency, among other parameters. For beams based on (\ref{eq:simpleCodebook})
or (\ref{eq:directionCodebook}), both $\theta_{\mathrm{max}}$ and
$\theta_{\mathrm{HPBW}}$ can be calculated and tabulated \cite{Orfandis2008}.\textcolor{red}{{} }

\subsubsection{Idealized Receiver}

For each transmitted beam, the receiver observes the following complex
baseband signal:
\begin{equation}
\mathbf{y}(t)=\sum_{k=0}^{K-1}\mathbf{H}_{k}\mathbf{f}_{m}x(t-\tau_{k})+\mathbf{n}(t),\label{receivedsignal}
\end{equation}
where $\mathbf{n}(t)\in\mathbb{C}^{N_{r}}$ is a Gaussian noise vector
with zero mean and two-sided power spectral density $N_{0}/2$. We
will consider an idealized receiver D2, which samples the entire signal
$\mathbf{y}(t)$ and is synchronized to D1 \cite{Shen2010}. While
such a receiver may be impractical, it allows us to understand the
ultimate performance of position-aided protocols and can thus serve
as a benchmark for different receiver structures with analog beamforming
as well as low-complexity algorithms. 

\subsection{Performance Metrics\label{subsec:Performance-Metrics}}

The beam selection protocol works in an iterative manner, where each
iteration $i\ge1$ involves selecting a number of active antennas
$N_{t}^{(i)}<N_{t}$ and a number of beams $M_{t}^{(i)}$. The objective
of protocol is to quickly determine a beamforming vector $\mathbf{f}_{\mathrm{sel}}$
resulting in high SNR. The relevant performance metrics are thus SNR,
number of transactions, and positioning quality. 
\begin{enumerate}
\item \textbf{SNR: }The selection of $\mathbf{f}_{\mathrm{sel}}$ intends
to maximize the SNR, defined as 
\begin{equation}
\mathrm{SNR}\triangleq\sum_{k=0}^{K-1}\frac{N_{t}N_{r}E_{s}}{N_{0}}|h_{k}|^{2}\left\Vert \mathbf{a}_{\mathrm{tx}}^{\mathrm{H}}(\theta_{\mathrm{tx},k})\mathbf{f}_{\mathrm{sel}}\right\Vert .\label{eq:SNR}
\end{equation}
\item \textbf{Number of transactions:} Considering an iterative beam selection
protocol, total beam selection time can be broken down for each iteration
$i\ge1$ as follows: (i) training stage during which D1 sends $M_{t}^{(i)}$
training sequences; (ii) feedback stage, during which D2 reports back
to D1; (iii) mapping stage, during which D1 informs D2 about the number
of required transmit beam patterns $M_{t}^{(i+1)}$; and a one time
(iv) acknowledgment after which high-rate data communication can start.
The total number of transactions $N_{\mathrm{trans}}$ can be quantified
as $N_{\mathrm{trans}}=\sum_{i=1}^{I}(M_{t}^{(i)}+2)+1$ . Note that
when the mapping is agreed a priori and feedback messages are neglected,
we find that 
\begin{equation}
N_{\mathrm{trans}}\approx\sum_{i=1}^{I}M_{t}^{(i)}.\label{eq:timeGeneral}
\end{equation}
\item \textbf{Positioning quality: }We consider the expected positioning
and orientation errors, given by
\begin{equation}
\mathbb{E}\{\Vert\mathbf{p}-\hat{\mathbf{p}}\Vert^{2}\}\label{eq:error_pos}
\end{equation}
and 
\begin{equation}
\mathbb{E}\{\Vert\alpha-\hat{\alpha}\Vert^{2}\},\label{eq:error_orient}
\end{equation}
where $\hat{\mathbf{p}}$ and $\hat{\alpha}$ denote the estimated
position and angle of rotation for the D2, respectively, obtained
from the sequence of received signals of the form (\ref{receivedsignal}). 
\end{enumerate}

\section{Conventional Beam Selection\label{sec:Conventional-Beam-Selection}}

A beam selection protocol with the goal of minimizing the beamforming
set-up time and mitigate the high path-loss has been adopted by the
IEEE 802.15.3c standard as an optional functionality \cite{Wang}.
This iterative protocol relies on a multi-level beam tree search starting
from lower resolution beams that cover large angular range per beam
moving towards higher resolution beams covering a smaller angular
range. \textcolor{black}{Other protocols have been considered in the
literature \cite{Lee2011,Chen2011,Xiao2016,Hur}}\textcolor{red}{{}
}. Here, we describe a general beam selection protocol for D1, not
exploiting or requiring any position information. 

\subsection{General Protocol Operation}

The iterative protocol selects a number of active antennas and a number
of beams at each iteration $i\ge1$. In particular, at iteration $i$,
D1 selects $M_{t}^{(i)}$ beams with associated beamforming vectors
\begin{equation}
\mathbf{F}^{(i)}=\{\mathbf{f}_{1}^{(i)},\ldots,\mathbf{f}_{M_{t}^{(i)}}^{(i)}\}
\end{equation}
to be used with $N_{t}^{(i)}\leq N_{t}$ selected active antennas.
The protocol makes use of a finite codebook from which beams can be
selected for each value of $N_{t}^{(i)}.$ D1 transmits a reference
signal $x(t)$ for each of the $M_{t}^{(i)}$ beams. Through suitable
signal processing, D2 measures the reference signal received power
(RSPS) $P_{m}^{(i)}$ for each of the $m=1,\ldots,M_{t}^{(i)}$ transmitted
beams, and gathers them in the vector $\mathbf{P}^{(i)}\in\mathbb{R}^{M_{t}^{(i)}}$
. The selection of the beams at each iteration $i$ is dependent on
the previous beam selection $\mathbf{F}^{(i-1)}$ and on the reference
signal received powers $\mathbf{P}^{(i-1)}$ transmitted as feedback
from D2 to D1. The mapping $\mathbf{F}^{(i)}=f_{\mathrm{map}}(\mathbf{F}^{(i-1)},\mathbf{P}^{(i-1)})$
depends on the specific codebook employed. The protocol is summarized
as pseudocode in Algorithm \ref{wangprotocol}. 
\begin{algorithm}[th]
\caption{\label{wangprotocol}RSPS protocol.}

\begin{algorithmic}[1]

\STATE \textbf{Input:} $N_{t}^{(1)}$, $M_{t}^{(1)}$ , $f_{\mathrm{map}}$,
and $N_{t}$

\WHILE{$N_{t}^{(i)} \leq N_{t}$}

\FOR{$m=1:M_{t}^{(i)}$}

\STATE D1 transmits a TS for each $m$-th beam pattern, $\mathbf{f}_{m}^{(i)}$;

\STATE D2 measures RSPS for each $m$-th pattern, $P_{m}^{(i)}$;

\ENDFOR

\STATE D2 transmits all measured received powers $\mathbf{P}^{(i)}$
to D1;

\STATE $i=i+1;$

\STATE A new selection of beams is obtained at D1: $\mathbf{F}^{(i)}=f_{\mathrm{map}}(\mathbf{F}^{(i-1)},\mathbf{P}^{(i-1)})$

\ENDWHILE

\STATE \textbf{Output: }Final beam pattern selection $\mathbf{f}_{\mathrm{sel}}$

\end{algorithmic}
\end{algorithm}

\subsection{Protocol-specific Performance \label{subsec:Realizations_snr}}

We now present the evaluation of the performance metrics in Section
\ref{subsec:Performance-Metrics} for the RSPS protocol.
\begin{enumerate}
\item \textbf{SNR: }Upon completion of the protocol, at iteration $I$,
corresponding to $N_{t}^{(I)}=N_{t},$ the beam with the highest RSPS
is selected
\begin{equation}
\mathbf{f}_{\mathrm{sel}}=\arg\max_{\mathbf{f}_{m}^{(I)}}P_{m}^{(I)}(\mathbf{f}_{m}^{(I)}),
\end{equation}
and the SNR is evaluated according to (\ref{eq:SNR}). 
\item \textbf{Number of transactions: }Depending on the codebook design,
which is both known at D1 and D2, different implementations of the
protocol can be designed by means of the mapping $\mathbf{F}^{(i)}=f_{\mathrm{map}}(\mathbf{F}^{(i-1)},\mathbf{P}^{(i-1)})$,
e.g, \cite{Garcia2016,Wang}. A simple mapping would involve one iteration
with $M_{t}=N_{t}$ narrow beams, leading to a number of transactions
$N_{\mathrm{trans}}^{\mathrm{conv}}\approx N_{t}$ \cite{Kim2014}.
A reduction in delay can be achieved through a multi-level beam search
from broad to directive beams, noting that the HPBW scales roughly
as $1/N_{t}^{(i)}$ \cite{Orfandis2008}, so that a beam with $N_{t}^{(i-1)}$
antennas can be covered with $M_{t}^{(i)}\le3$ beams with $N_{t}^{(i)}=2N_{t}^{(i-1)}$
antennas. This leads to $I=\log_{2}(N_{t})$ and thus $N_{\mathrm{trans}}^{\mathrm{conv}}\approx3\log_{2}(N_{t})$.
\item \textbf{Positioning quality: }The protocol does not provide any positioning
information. 
\end{enumerate}

\section{Proposed Joint Positioning and Beam Selection\label{sec:Joint-PBS}}

In this section, we introduce the proposed iterative position-based
beam selection protocol. The protocol aims to minimize the set-up
time and mitigate the high path-loss using D2 position information
as proxy for the optimal pointing of the transmit beams. Before we
describe the proposed protocol, followed by its performance and implementation
details, we first briefly detail some properties of mm-wave positioning. 

\subsection{Performance of Mm-wave Positioning}

D2 can perform estimation of its position and orientation (represented
by $\bm{\beta}$) based on the received waveforms from D1. The quality
of such estimation can be assessed through the Fisher information
matrix (FIM) \cite{Kay}. Given the statistics of a waveform $\mathbf{y}(t)$
of the form (\ref{receivedsignal}) and an unknown vector parameter
\[
\bm{\eta}=\left[\tau_{0},\mathbf{\bm{\theta}}_{0}^{\mathrm{T}},\mathbf{h}_{0}^{\mathrm{T}},\ldots,\tau_{K-1},\mathbf{\bm{\theta}}_{K-1}^{\mathrm{T}},\mathbf{h}_{K-1}^{\mathrm{T}}\right]^{\mathrm{T}},
\]
where $\bm{\theta}_{k}=\left[\theta_{\mathrm{tx},k},\theta_{\mathrm{rx},k}\right]^{\mathrm{T}}$,
$\mathbf{h}_{k}=\left[h_{R,k},h_{I,k}\right]=[\Re\{h_{k}\},\Im\{h_{k}\}]$,
the FIM associated with a single beam, $\mathbf{J}_{\bm{\mathbf{\eta}}}^{(\mathrm{beam})}$,
is a $5K\times5K$ matrix, whose expression is provided in (\ref{FIM_gral})
in the Appendix. 

While each $\mathbf{y}(t)$ corresponds to a single transmit beam,
the FIM for multiple beams is simply the sum of the corresponding
FIMs, due to the additive nature of Fisher information. For each iteration
$i$ of a beam selection protocol, we can thus compute the FIM associated
with the $m$-th beam, say, $\mathbf{J}_{\bm{\eta}}^{(i,m)}$. The
total FIM after $i$ iterations can then be expressed as

\begin{equation}
\mathbf{J}_{\bm{\eta}}^{(i)}=\sum_{l=1}^{i}\sum_{m=1}^{M_{t}^{(l)}}\mathbf{J}_{\bm{\eta}}^{(l,m)}.
\end{equation}
Since there is an injective relation between $\bm{\eta}$ and\footnote{In a practical implementation a priori information on the number of
scatterers is not required \cite{Shahmansoori2017}. } 
\[
\bm{\eta}'=\left[\bm{\beta}^{\mathrm{T}},\mathbf{h}_{0}^{\mathrm{T}},\mathbf{s}_{1}^{\mathrm{T}},\mathbf{h}_{1}^{\mathrm{T}},\ldots,\mathbf{s}_{K-1}^{\mathrm{T}},\mathbf{h}_{K-1}^{\mathrm{T}}\right]^{\mathrm{T}},
\]
we can also determine the FIM of $\mathbf{J}_{\bm{\eta}'}^{(i)}$
as $\mathbf{J}_{\bm{\eta}'}^{(i)}=\mathbf{T}^{\mathrm{T}}\mathbf{J}_{\bm{\eta}}^{(i)}\mathbf{T}$,
where $\mathbf{T}$ is the Jacobian matrix associated with the transformation
from $\bm{\eta}$ to $\bm{\eta}'$, that is, $T_{ij}=\partial\eta_{i}/\partial\eta'_{j}$
.

Finally, the inverse of the FIM can be related to the mean squared
error (MSE) of unbiased estimators of $\bm{\eta}'$ \cite{Kay}: 
\begin{equation}
\mathbb{E}_{\mathbf{y}|\bm{\eta}'}\left[\left(\hat{\bm{\eta}}'-\bm{\text{\ensuremath{\eta}}}\right)\left(\hat{\bm{\eta}}'-\bm{\text{\ensuremath{\eta}}}\right)^{\mathrm{T}}\right]\succeq\left[\mathbf{J}_{\bm{\eta}'}^{(i)}\right]^{-1}.
\end{equation}
From this relationship, we can immediately derive the so-called position
error bound (PEB) and rotation error bound (REB) as
\begin{align}
\mathrm{PEB}^{(i)} & =\sqrt{\mathrm{tr}\left\{ \left[\mathbf{J}_{\bm{\eta}'}^{(i)}\right]_{1:2,1:2}^{-1}\right\} }\label{eq:PEBdef}\\
 & \le\sqrt{\mathbb{E}\{\Vert\mathbf{p}-\hat{\mathbf{p}}\Vert^{2}\}}
\end{align}
and 
\begin{align}
\mathrm{REB}^{(i)} & =\sqrt{\left[\mathbf{J}_{\bm{\eta}'}^{(i)}\right]_{3,3}^{-1}}\label{eq:REBdef}\\
 & \le\sqrt{\mathbb{E}\{\Vert\alpha-\hat{\alpha}\Vert^{2}\}},
\end{align}
where $\left[\cdot\right]_{1:2,1:2}^{-1}$ denotes the $2\times2$
upper left submatrix of the inverse of the argument, and $\left[\cdot\right]_{3,3}^{-1}$
denotes the third diagonal element of the inverse of the argument. 

\subsubsection*{Remarks }
\begin{itemize}
\item We note that in contrast to conventional range-based positioning,
the use of multiple antennas at both devices allows for the determination
of both the position and the orientation of D2 using signals from
a single reference device D1. 
\item The FIM and corresponding PEB and REB are valid, irrespective of the
subsequent processing at the receiver. Such processing can include
analog beamforming as well as low-complexity estimation and detection
algorithms.
\item Given the geometric model, the FIM does not account for path resolvability
in time and angle spaces. However, this can be addressed as follows.
Consider paths $a$ and $b$, with AOAs $\theta_{r,a}$, and $\theta_{r,b}$;
delays $\tau_{a},$ and $\tau_{b}$, respectively. Paths $a$ and
$b$ are considered unresolvable in time \emph{and} angle, when $\left|\tau_{a}-\tau_{b}\right|\le1/B$
and $N_{r}\lambda\left|\sin(\theta_{r,a})-\sin(\theta_{r,b})\right|\le d$
\textcolor{black}{\cite{Tse2005}}. When two paths are unresolvable,
they are to be combined into a single path by adding the complex channel
gains, prior to computation of the Fisher information. 
\end{itemize}

\subsection{General Protocol Operation\label{subsec:Positioning-based-algorithm}}

From the above FIM analysis, it is apparent that D2 can not only compute
the received powers for each beam, but also harness them to compute
its position and orientation $\bm{\beta}=[\mathbf{p},\alpha]^{\mathrm{T}}$.
We will denote the aggregated waveforms at iteration $i$ by $\mathbf{y}^{(i)}$,
and the collection of $\mathbf{y}^{(i)}$ up to iteration $i$ by
$\mathbf{y}^{(1:i)}$ . Our idealized receiver D2 can thus be equipped
with an estimator, which can determine an estimate of $\bm{\beta}$
from $\mathbf{y}^{(1:i)}$ at the end of each iteration $i$, operating
close to the fundamental performance bounds (\ref{eq:PEBdef})\textendash (\ref{eq:REBdef}).
Considering a Gaussian approximation of the position and orientation
error, we can describe the estimate by a mean $\hat{\bm{\beta}}^{(i)}$
and a covariance matrix $\bm{\Sigma}_{\bm{\beta}}^{(i)}$. This information
can be fed back to D1. The protocol then operates according to Algorithm
\ref{alg:positionprotocol-simplified}. Since D1 has more information
about D2 than in the conventional algorithm, a more intelligent mapping
function can be designed, as will be described in Section \ref{subsec:Realizations_pos}.
In addition, both D2 and D1 have knowledge of D2's position and orientation. 

\begin{algorithm}[h]
\caption{Joint positioning and beam selection protocol.\label{alg:positionprotocol-simplified}}

\begin{algorithmic}[1]

\STATE \textbf{Input:} $N_{t}^{(1)}$, $M_{t}^{(1)}$ , $f_{\mathrm{map}}^{\mathrm{pos}}$
and $N_{t}$

\WHILE{$N_{t}^{(i)} \leq N_{t}$}

\FOR{$m=1:M_{t}^{(i)}$}

\STATE D1 transmits a TS for each $m$-th beam pattern, $\mathbf{f}_{m}^{(i)}$;

\STATE D2 measures received power for each $m$-th beam pattern,
$P_{m}^{(i)}$;

\ENDFOR

\STATE D2 determines $[\hat{\bm{\beta}}^{(i)},\bm{\Sigma}_{\bm{\beta}}^{(i)}]$;

\STATE D2 feeds back $[\mathbf{P}^{(i)},\hat{\bm{\beta}}^{(i)},\bm{\Sigma}_{\bm{\beta}}^{(i)}]$
to D1;

\STATE $i=i+1;$

\STATE A new selection of beams is obtained at D1: $\mathbf{F}^{(i)}=f_{\mathrm{map}}^{\mathrm{pos}}(\mathbf{F}^{(i-1)},\mathbf{P}^{(i)},\hat{\bm{\beta}}^{(i)},\bm{\Sigma}_{\bm{\beta}}^{(i)})$;

\ENDWHILE

\STATE \textbf{Output: }Final beam pattern selection $\mathbf{f}_{\mathrm{sel}}$,
final D2's position and orientation $\hat{\bm{\beta}}^{(i)}$ and
their uncertainties $\bm{\Sigma}_{\bm{\beta}}^{(i)}$.

\end{algorithmic}
\end{algorithm}

\subsection{Protocol-specific Performance \label{subsec:Realizations_pos}}

We now present the evaluation of the performance metrics in Section
\ref{subsec:Performance-Metrics} for the protocol described by Algorithm
\ref{alg:positionprotocol-simplified}.
\begin{enumerate}
\item \textbf{SNR: }As for the conventional protocol, the SNR is computed
using (\ref{eq:SNR}), based on the final selected beam.
\item \textbf{Number of transactions: }Different implementations depending
on the mapping function $f_{\mathrm{map}}^{\mathrm{pos}}(\mathbf{F}^{(i-1)},\mathbf{P}^{(i)},\hat{\bm{\beta}}^{(i)},\bm{\Sigma}_{\bm{\beta}}^{(i)})$
can be designed, in order to reduce the number of transactions. In
particular, when D1 has knowledge of the AOD, it can select an appropriate
number of active antennas and beams. More specifically, D1 can process
$\hat{\bm{\beta}}^{(i)},\bm{\Sigma}_{\bm{\beta}}^{(i)}$ to compute
an AOD estimate $\hat{\theta}_{\mathrm{tx},0}^{(i)}$ and the AOD
standard deviation, denoted as $\sigma_{\mathrm{tx},0}^{(i)}$. A
conventional hierarchical protocol with $N_{t}^{(i)}=2N_{t}^{(i-1)}$
and $M_{t}^{(i)}=3$ can be used whenever the AOD uncertainty is large,
i.e., $\sigma_{\mathrm{tx},0}^{(i)}\ge3\theta_{\mathrm{HPBW}}(2N_{t}^{(i-1)},\hat{\theta}_{\mathrm{tx},0}^{(i)})$.
On the other hand, if $\sigma_{\mathrm{tx},0}^{(i)}$$<$$3\theta_{\mathrm{HPBW}}(2N_{t}^{(i-1)},\hat{\theta}_{\mathrm{tx},0}^{(i)})$,
the number of transactions can be reduced by using more than $2N_{t}^{(i)}$
active antennas with $M_{t}^{(i)}=3$. In the latter case, we propose
to set $N_{t}^{(i)}$ according to
\begin{align}
\mathrm{maximize}\,\,\, & N_{t}^{(i)}\\
\mathrm{subject\,to}\,\,\, & N_{t}^{(i)}\le N_{t}\\
 & 3\theta_{\mathrm{HPBW}}(N_{t}^{(i)},\hat{\theta}_{\mathrm{tx},0}^{(i)})\ge\sigma_{\mathrm{tx},0}^{(i)},\label{eq:AODcondition}
\end{align}
and transmit 3 beams covering the AOD region $\hat{\theta}_{\mathrm{tx},0}^{(i)}\pm\sigma_{\mathrm{tx},0}^{(i)}$.
Such beams can also be optimized to minimize the future expected uncertainty,
as detailed in Appendix \ref{sec:Optimized-Beam-Directions}. We thus
expect that in cases when $\bm{\beta}$ can be accurately estimated
with few antennas, then the number of transactions fulfills $N_{\mathrm{trans}}^{\mathrm{pos}}\ll N_{\mathrm{trans}}^{\mathrm{conv}}$,
and $N_{\mathrm{trans}}^{\mathrm{pos}}\approx N_{\mathrm{trans}}^{\mathrm{conv}}$
otherwise. From \cite{Shamansoori2015,Dardari}, it is known that
good estimates of $\bm{\beta}$ are possible when enough beams are
transmitted pointing roughly in the direction of D2, and the received
SNR associated with those beams is sufficiently high. Consequently,
we expect $N_{\mathrm{trans}}^{\mathrm{pos}}\ll N_{\mathrm{trans}}^{\mathrm{conv}}$
for D2 locations close to D1.  
\item \textbf{Positioning quality: }The proposed protocol can be assessed
in terms of position (\ref{eq:error_pos}) and orientation errors
(\ref{eq:error_orient}), which can be predicted through the FIM. 
\end{enumerate}

\begin{figure}
\begin{centering}
\includegraphics[width=1\columnwidth]{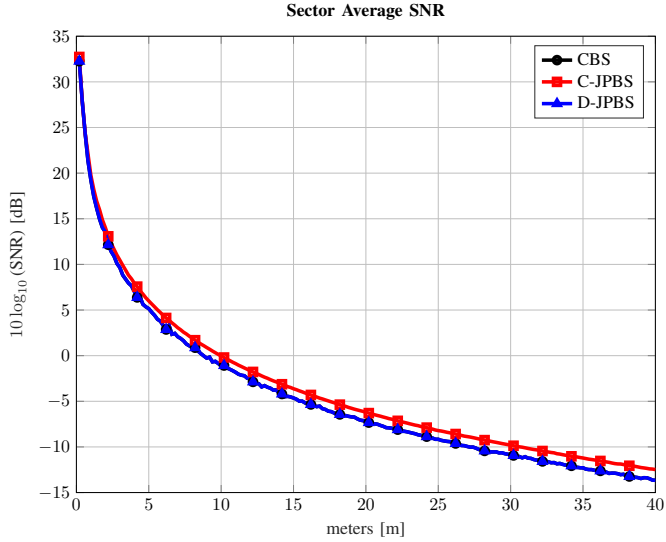}
\par\end{centering}
\caption{Average SNR as a function of distance to D2. \label{fig:SNRsector}}
\end{figure}

\section{Simulation Results\label{sec:Simulation-Results}}

\subsection{Simulation Setup}

We consider a $40\,\mathrm{m}\times40\,\mathrm{m}$ area where the
receiver D2 can be placed, D1 is located at a fixed and known position
$\mathbf{q}=[0,0]^{\mathrm{T}}$, and a scatterer is fixed at $\mathbf{s}=[5,5]^{\mathrm{T}}$,
hence $K=2$ . We set $f_{c}=60$ GHz, $B=100$ MHz, $N_{0}=-84$
dBm/GHz, $\alpha=0$ rad. For the LOS path, we set $h_{0}=\exp(-j2\pi f_{c}\tau_{0})/\sqrt{\rho_{0}}$,
where $\rho_{0}=(2\pi\Vert\mathbf{q}-\mathbf{p}\Vert/\lambda)^{2}$
is the path-loss between D1 and D2. For the NLOS paths, we set $h_{k}=\exp(-j2\pi f_{c}\tau_{k})/\sqrt{\rho_{k}}$,
in which $\rho_{k}=(2\pi(\Vert\mathbf{q}-\mathbf{s}_{k}\Vert\times\Vert\mathbf{s}_{k}-\mathbf{p}\Vert)/\lambda)^{2}$
\cite{Rappaport}. The number of antennas at both D1 and D2 is $N_{t}=N_{r}=64$,
and the inter-element spacing is $d=\lambda/2$. The ULAs are located
along the vertical axis. We generate a signal $x(t)$ with $N=64$
symbols. We set remaining parameters such that the SNR given by (\ref{eq:SNR})
on the horizontal axis at 10 meters from D1 is 0 dB (i.e., the nominal
communication range is 10 m). 

We will evaluate three protocols in terms of SNR, number of transactions,
and positioning quality (using $\mathrm{PEB}^{(I)}$ and $\mathrm{REB}^{(I)}$):

\begin{itemize}
\item A conventional beam selection protocol (termed CBS), similar to \cite{Wang},
which can use only $M_{t}\in\{2,4,8,16,32,64\}$ discrete beams and
$N'_{t}=M_{t}$ active contiguous antennas sequentially selected,
generated using phase shifters $\phi_{i}=\{\pi,-\pi,\pi/2,-\pi/2\}$. 
\item A discretized joint positioning and beam selection protocol (termed
D-JPBS), using the same discrete codebook as the conventional beam
selection protocol. 
\item A joint positioning and beam selection protocol (termed C-JPBS), with
a continuous codebook of the form (\ref{eq:directionCodebook}), where
$\theta_{m}\in[-\pi/2,\pi/2]$ and $N'_{t}\in[2,64]$. 
\end{itemize}
\begin{figure}
\begin{centering}
\includegraphics[width=1\columnwidth]{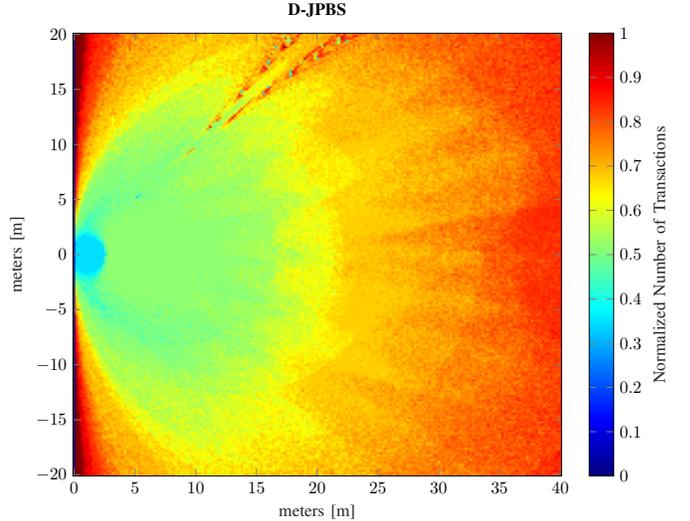}
\par\end{centering}
\caption{Normalized number of transactions with respect to the conventional
beam selection protocol for the discrete joint positioning and beam
selection protocol.\label{fig:Transactionsdjp}}
\end{figure}

\subsection{Results and Discussion}

We show results in two forms: contour plots over the area showing
average performance (over 30 realizations) for the D-JPBS and plots
that show the performance as a function of distance, where we averaged
the values of the contour plots along concentric circles around D1. 

\subsubsection{Final SNR}

Figure \ref{fig:SNRsector} shows the SNR as a function of the D1-D2
distance for the CBS, C-JPBS, and D-JPBS, respectively. Overall, all
protocols show a similar performance in terms of SNR. The CBS and
D-JPBS protocols show identical performance. We can observe that the
C-JPBS protocol achieves a slightly higher SNR with increasing distance
between devices. The higher SNR is due to the higher degree of freedom
that the C-JPBS protocol has compared to the other two protocols,
and which allows the C-JPBS protocol to point the beams directly to
the position of D2. In contrast, the CBS and D-JPBS protocols employ
a more restricted codebook, and hence the SNR is dependent on the
discretization of the beams. We conclude that the positioning-based
protocols have no significant negative impact on the final SNR. 

\subsubsection{Number of transactions}

\begin{figure}
\begin{centering}
\includegraphics[width=1\columnwidth]{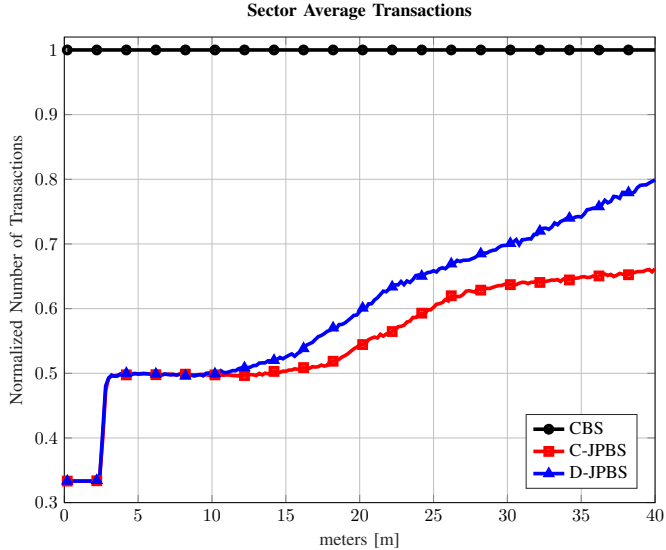}
\par\end{centering}
\caption{Average number of normalized transactions with respect to the conventional
beam selection protocol as a function of distance to D1.\label{fig:Transsector}}
\end{figure}

The CBS uses a fixed number of transactions to complete the procedure
regardless of D2 location. Hence, we show the contour plot of the
\emph{normalized} number of transactions with respect to the fixed
CBS transactions in Figure~\ref{fig:Transactionsdjp} for the D-JPBS
protocol, as an example of the behavior of the protocol in terms of
transactions. It can be observed that number of transactions is dependent
on the discretization of the beams. We note distinct regions in the
figure due to discrete number of antennas that can be used by the
protocol, combined with the criterion (\ref{eq:AODcondition}). Since
the beams are wider at the endfire of the D1 array, more transactions
are used in the upper and lower left regions of the areas. Moreover,
behind the scatterer we can observe a peculiar behavior caused by
the inability of D1 in such locations to estimate both its own location
and the scatterer location. In particular, the paths within this region
are resolvable in angle but not in delay creating the need for more
transactions. Figure~\ref{fig:Transsector} shows the number of transactions
as a function of D1-D2 distance. As distance grows larger, the number
of transactions for both the D-JPBS and C-JPBS protocols increases,
since we need more information in the FIM to jump to a higher number
of contiguous active antennas. We can observe a reduction of $67\%$
in the number of transactions is achieved when D2 is close to D1 (3
meters or less). The reduction grows to $50\%$ at inter-device distances
between 3 and 10 meters. Beyond 10 meters we start observing a gap
between the D-JPBS and C-JPBS protocols. This is due to the codebook
restriction in the D-JPBS. The beam discretization has more influence
at larger distances, given the separation between the beams; thus,
giving an advantage to the C-JPBS protocol which has no codebook restrictions
and beams can be pointed at any direction. We can conclude that position
information has an impact in the reduction of latency of the device-to-device
beam selection protocol.

\subsubsection{Positioning performance}

\begin{figure}
\begin{centering}
\includegraphics[width=1\columnwidth]{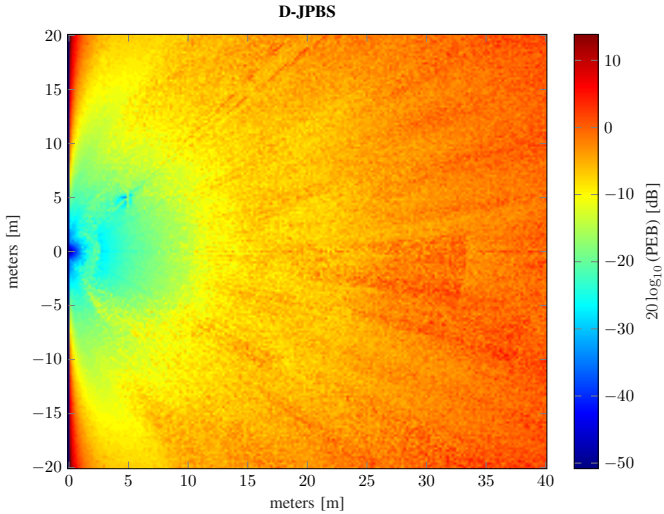}
\par\end{centering}
\caption{PEB for the discrete joint positioning and beam selection protocol.\label{fig:PEBdpos}}
\end{figure}
\begin{figure}
\begin{centering}
\includegraphics[width=1\columnwidth]{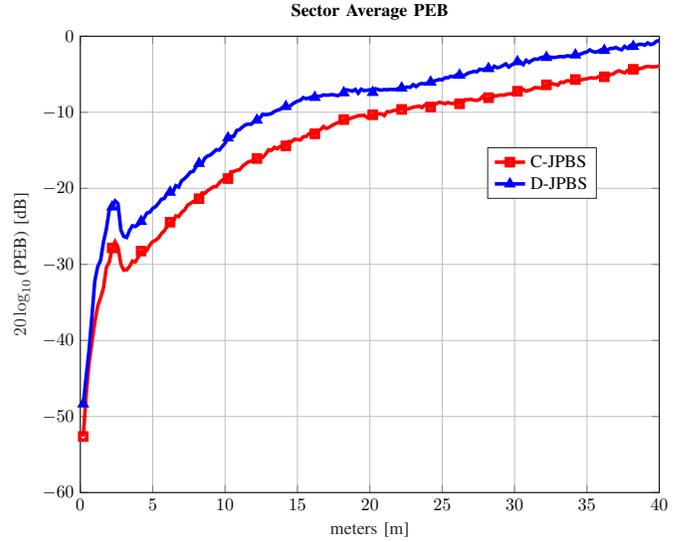}
\par\end{centering}
\caption{Average PEB as a function of distance to D1.\label{fig:PEBsector}}
\end{figure}

Figure \ref{fig:PEBdpos} shows the PEB for the D-JPBS. Note that
a PEB of 0 dB corresponds to an uncertainty of 1 m, 10 dB corresponds
to 3.2 m, and -10 dB to 30 cm. The achieved PEB values depend on the
choice of the number of symbols $N$. As expected, the PEB values
also depend on the distance and AOD with respect to D2. Due to high
SNR gains, very low PEB values are observed close to D1. Moreover,
in the region around the position of the scatterer $\mathbf{s}=[5,5]^{\mathrm{T}}$,
we can observe a small decrease in PEB due to the aggregate information
to the FIM provided by the scatterer. This behavior is only apparent
close to the scatterer given our path loss model for the scattered
path. The discretized protocol shows more accurate PEB in the directions
of the available beams within the codebook. We note that within the
region behind the scatterer there is a decrease in PEB due to the
poor resolvability of the paths and thus the inability of D2 to estimate
both its own location and the scatterer location. The paths for this
region are not resolvable in time, but are resolvable in the angle
domain. This translates into poor delay information, which causes
a degradation of the estimated parameters and hence of the PEB. 

Figure \ref{fig:PEBsector} shows the PEB as a function the D1-D2
distance. The general trend is that the PEB increases as a function
of distance, but we can observe a change around 3 meters, where the
PEB abruptly decreases due to the increase in number of transactions,
which provide more information to the FIM. Similar behavior is observed
for the REB, hence only the PEB figures are presented. We can observe
how accurate position information is attainable within a moderate
distance between devices.

\section{Conclusions\label{sec:Conclusions}}

Beam selection is an essential step in establishing a mm-wave communication
link. Conventional protocols rely on measuring the received power
obtained with a set of beamformers, which are successively made more
and more directive. Given the characteristics of mm-wave propagation
and the use of standard geometric channel models, we have exploited
the ability of devices to determine their location during the beam
selection process and thus improve the subsequent selection of beams.
We have shown that such in-band position-aided protocols have similar
performance as the conventional protocol in terms of achieved final
SNR, but they are significantly faster and can additionally provide
the position or orientation of the device in an accurate manner. Such
information can be used in other procedures or applications. Our analysis
indicates that standard codebooks can be used to harness these gains,
with similar performance to more complex codebooks. Future work will
include the removal of assumptions in the idealized receiver, such
as the introduction of receiver beamforming.

\appendices{}

\section{Derivation of the FIM}

\begin{figure*}
\begin{equation}
\mathbf{J}_{(i,j)}=\left[\begin{array}{ccccc}
\Phi(\tau_{i},\tau_{j}) & \Phi(\tau_{i},\theta_{\mathrm{tx},j}) & \Phi(\tau_{i},\theta_{\mathrm{rx},j}) & \Phi(\tau_{i},h_{R,j}) & \Phi(\tau_{i},h_{I,j})\\
\Phi(\theta_{\mathrm{tx},i},\tau_{j}) & \Phi(\theta_{\mathrm{tx},i},\theta_{\mathrm{tx},j}) & \Phi(\theta_{\mathrm{tx},i},\theta_{\mathrm{rx},j}) & \Phi(\theta_{\mathrm{tx},i},h_{R,j}) & \Phi(\theta_{\mathrm{tx},i},h_{I,j})\\
\Phi(\theta_{\mathrm{rx},i},\tau_{j}) & \Phi(\theta_{\mathrm{rx},i},\theta_{\mathrm{tx},j}) & \Phi(\theta_{\mathrm{rx},i},\theta_{\mathrm{rx},j}) & \Phi(\theta_{\mathrm{rx},i},h_{R,j}) & \Phi(\theta_{\mathrm{rx},i},h_{I,j})\\
\Phi(h_{R,i},\tau_{j}) & \Phi(h_{R,i},\theta_{\mathrm{tx},j}) & \Phi(h_{R,i},\theta_{\mathrm{rx},j}) & \Phi(h_{R,i},h_{R,j}) & \Phi(h_{R,i},h_{I,j})\\
\Phi(h_{I,i},\tau_{j}) & \Phi(h_{I,i},\theta_{\mathrm{tx},j}) & \Phi(h_{I,i},\theta_{\mathrm{rx},j}) & \Phi(h_{I,i},h_{R,j}) & \Phi(h_{I,i},h_{I,j})
\end{array}\right]\label{eq:FIMsub}
\end{equation}
\end{figure*}
We consider the case of multiple paths and a single beam. The general
form of the FIM for $k$ paths is given by

\begin{equation}
\mathbf{J}_{\bm{\mathbf{\eta}}}^{(\mathrm{beam})}=\left[\begin{array}{ccc}
\mathbf{J}_{(1,k)} & \cdots & \mathbf{J}_{(1,k)}\\
\vdots & \ddots & \vdots\\
\mathbf{J}_{(k,1)} & \cdots & \mathbf{J}_{(k,k)}
\end{array}\right],\label{FIM_gral}
\end{equation}
where each of the sub-matrices has the form (\ref{eq:FIMsub}), in
which 
\begin{equation}
\Phi(x_{1},x_{2})=\mathbb{E}_{\mathbf{y},\mathbf{a}|\bm{\eta}}\left\{ \frac{\partial}{\partial x_{1}}\Lambda(\mathbf{y}|\bm{\eta},\mathbf{a})\left(\frac{\partial}{\partial x_{2}}\Lambda(\mathbf{y}|\bm{\eta},\mathbf{a})\right)^{*}\right\} ,
\end{equation}
where the log-likelihood function is expressed as
\begin{equation}
\Lambda(\mathbf{y}|\bm{\eta},\mathbf{a})=-\frac{1}{N_{0}}\int\Bigl\Vert\mathbf{y}(t)-\sum_{k=0}^{K-1}\mathbf{H}_{k}\mathbf{f}x(t-\tau_{k})\Bigr\Vert^{2}\mathrm{d}t.
\end{equation}
Denoting the noise-free signal by

\begin{equation}
\mathbf{m}(t)=\sum_{k=0}^{K-1}\mathbf{H}_{k}\mathbf{f}x(t-\tau_{k}),
\end{equation}
it can be shown that 
\begin{align}
\Phi(x_{1},x_{2}) & =\frac{2}{N_{0}}\Re\left\{ \mathbb{E}_{\mathbf{a}}\left\{ \int\frac{\partial\mathbf{m}^{\mathrm{H}}(t)}{\partial x_{1}}\frac{\partial\mathbf{m}(t)}{\partial x_{2}}\mathrm{d}t\right\} \right\} .
\end{align}
It is readily verified that for an arbitrary path $i$
\begin{align}
\frac{\partial\mathbf{m}(t)}{\partial\tau_{i}} & =-\sqrt{N_{t}N_{r}}h_{i}\mathbf{a}_{\mathrm{rx}}(\theta_{\mathrm{rx},i})\mathbf{a}_{\mathrm{tx}}^{\mathrm{H}}(\theta_{\mathrm{tx},i})\mathbf{f}\dot{x}(t-\tau_{i})\\
\frac{\partial\mathbf{m}(t)}{\partial\theta_{\mathrm{tx},i}} & =\sqrt{N_{t}N_{r}}h_{i}\mathbf{a}_{\mathrm{rx}}(\theta_{\mathrm{rx},i})\mathbf{\dot{a}}_{\mathrm{tx}}^{\mathrm{H}}(\theta_{\mathrm{tx},i})\mathbf{f}x(t-\tau_{i})\\
\frac{\partial\mathbf{m}(t)}{\partial\theta_{\mathrm{rx},i}} & =\sqrt{N_{t}N_{r}}h_{i}\dot{\mathbf{a}}_{\mathrm{rx}}(\theta_{\mathrm{rx},i})\mathbf{a}_{\mathrm{tx}}^{\mathrm{H}}(\theta_{\mathrm{tx},i})\mathbf{f}x(t-\tau_{i})\\
\frac{\partial\mathbf{m}(t)}{\partial h_{R,i}} & =\sqrt{N_{t}N_{r}}\mathbf{a}_{\mathrm{rx}}(\theta_{\mathrm{rx},i})\mathbf{a}_{\mathrm{tx}}^{\mathrm{H}}(\theta_{\mathrm{tx},i})\mathbf{f}x(t-\tau_{i})\\
\frac{\partial\mathbf{m}(t)}{\partial h_{I,i}} & =\sqrt{N_{t}N_{r}}j\mathbf{a}_{\mathrm{rx}}(\theta_{\mathrm{rx},i})\mathbf{a}_{\mathrm{tx}}^{\mathrm{H}}(\theta_{\mathrm{tx},i})\mathbf{f}x(t-\tau_{i}).
\end{align}

\subsection*{Diagonal elements of the FIM}

We easily find that the diagonal elements of the submatrices $\mathbf{J}_{(i,j)}$.
We first define $1/\sigma^{2}=2NE_{s}N_{r}N_{t}/N_{0}$, $\gamma_{\mathrm{tx},i}=\mathbf{f}^{\mathrm{H}}\mathbf{a}_{\mathrm{tx}}(\theta_{\mathrm{tx},i})$,
$\beta_{ij}=\mathbf{a}_{\mathrm{rx}}^{\mathrm{H}}(\theta_{\mathrm{rx},i})\mathbf{a}_{\mathrm{rx}}(\theta_{\mathrm{rx},j})$
as well as $\dot{\mathbf{a}}_{\mathrm{tx}}(\theta_{\mathrm{tx}})=\partial\mathbf{a}_{\mathrm{tx}}(\theta_{\mathrm{tx}})/\partial\theta_{\mathrm{tx}}$
, $\dot{\gamma}_{\mathrm{tx,i}}=\dot{\mathbf{a}}_{\mathrm{tx}}^{\mathrm{H}}(\theta_{\mathrm{tx},i})\mathbf{f}$,
and $\ddot{\beta}_{ij}=\dot{\mathbf{a}}_{\mathrm{rx}}^{\mathrm{H}}(\theta_{\mathrm{rx},i})\dot{\mathbf{a}}_{\mathrm{rx}}(\theta_{\mathrm{rx},j})$.
We also introduce 
\begin{align*}
A_{0}(\Delta) & =\int p^{*}(t-\Delta)p(t)\mathrm{d}t\\
A_{1}(\Delta) & =\int\dot{p}^{*}(t-\Delta)p(t)\mathrm{d}t\\
A_{2}(\Delta) & =\int\dot{p}^{*}(t-\Delta)\dot{p}(t)\mathrm{d}t.
\end{align*}
We then find that 
\begin{align*}
\Phi(\tau_{i},\tau_{j}) & =\frac{1}{\sigma^{2}}\Re\left\{ h_{i}^{*}h_{j}\gamma_{\mathrm{tx},i}\gamma_{\mathrm{tx},j}^{*}\beta_{ij}A_{2}(\Delta_{ij})\right\} ,\\
\Phi(\theta_{\mathrm{tx},i},\theta_{\mathrm{tx},j}) & =\frac{1}{\sigma^{2}}\Re\left\{ h_{i}^{*}h_{j}\dot{\gamma}_{\mathrm{tx},i}^{*}\beta_{ij}\dot{\gamma}_{\mathrm{tx},j}A_{0}(\Delta_{ij})\right\} \\
\Phi(\theta_{\mathrm{rx},i},\theta_{\mathrm{rx},j}) & =\frac{1}{\sigma^{2}}\Re\left\{ h_{i}^{*}h_{j}\gamma_{\mathrm{tx},i}\ddot{\beta}_{ij}\gamma_{\mathrm{tx},j}^{*}A_{0}(\Delta_{ij})\right\} \\
\Phi(h_{R,i},h_{R,j}) & =\Phi(h_{I,i},h_{I,j})=\frac{1}{\sigma^{2}}\Re\left\{ \gamma_{\mathrm{tx},i}\gamma_{\mathrm{tx},j}^{*}\beta_{ij}A_{0}(\Delta_{ij})\right\} 
\end{align*}
where $\Delta_{ij}=\tau_{i}-\tau_{j}$. 

\subsection*{Off-diagonal elements of the FIM}

The off-diagonal elements are computed in similar fashion. Introducing,
$\dot{\beta}_{ij}=\mathbf{a}_{\mathrm{rx}}^{\mathrm{H}}(\theta_{\mathrm{rx},i})\dot{\mathbf{a}}_{\mathrm{rx}}(\theta_{\mathrm{rx},j})$,
the final expressions for the upper diagonal elements are computed
as:
\begin{align*}
\Phi(\tau_{i},\theta_{\mathrm{tx},j}) & =-\frac{1}{\sigma^{2}}\Re\left\{ h_{i}^{*}h_{j}\gamma_{\mathrm{tx},i}\beta_{ij}\dot{\gamma}_{\mathrm{tx},j}A_{1}(\Delta_{ij})\right\} \\
\Phi(\tau_{i},\theta_{\mathrm{rx},j}) & =-\frac{1}{\sigma^{2}}\Re\left\{ h_{i}^{*}h_{j}\gamma_{\mathrm{tx},i}\dot{\beta}_{ij}\gamma_{\mathrm{tx},j}^{*}A_{1}(\Delta_{ij})\right\} \\
\Phi(\tau_{i},h_{R,j}) & =-\frac{1}{\sigma^{2}}\Re\left\{ h_{i}^{*}\gamma_{\mathrm{tx},i}\beta_{ij}\gamma_{\mathrm{tx},j}^{*}A_{1}(\Delta_{ij})\right\} \\
\Phi(\tau_{i},h_{I,j}) & =-\frac{1}{\sigma^{2}}\Re\left\{ jh_{i}^{*}\gamma_{\mathrm{tx},i}\beta_{ij}\gamma_{\mathrm{tx},j}^{*}A_{1}(\Delta_{ij})\right\} \\
\Phi(\theta_{\mathrm{tx},i},\theta_{\mathrm{rx},j}) & =\frac{1}{\sigma^{2}}\Re\left\{ h_{i}^{*}h_{j}\dot{\gamma}_{\mathrm{tx},i}^{*}\dot{\beta}_{ij}\gamma_{\mathrm{tx},j}^{*}A_{0}(\Delta_{ij})\right\} \\
\Phi(\theta_{\mathrm{tx},i},h_{R,j}) & =\frac{1}{\sigma^{2}}\Re\left\{ h_{i}^{*}\dot{\gamma}_{\mathrm{tx},i}^{*}\beta_{ij}\gamma_{\mathrm{tx},j}^{*}A_{0}(\Delta_{ij})\right\} \\
\Phi(\theta_{\mathrm{tx},i},h_{I,j}) & =\frac{1}{\sigma^{2}}\Re\left\{ jh_{i}^{*}\dot{\gamma}_{\mathrm{tx},i}^{*}\beta_{ij}\gamma_{\mathrm{tx},j}^{*}A_{0}(\Delta_{ij})\right\} \\
\Phi(\theta_{\mathrm{rx},i},h_{R,j}) & =\frac{1}{\sigma^{2}}\Re\left\{ h_{i}^{*}\gamma_{\mathrm{tx},i}\dot{\beta}_{ji}^{*}\gamma_{\mathrm{tx},j}^{*}A_{0}(\Delta_{ij})\right\} \\
\Phi(\theta_{\mathrm{rx},i},h_{I,j}) & =\frac{1}{\sigma^{2}}\Re\left\{ jh_{i}^{*}\gamma_{\mathrm{tx},i}\dot{\beta}_{ij}^{*}\gamma_{\mathrm{tx},j}^{*}A_{0}(\Delta_{ij})\right\} \\
\Phi(h_{R,i},h_{I,i}) & =\frac{1}{\sigma^{2}}\Re\left\{ j\gamma_{\mathrm{tx},i}\beta_{ij}\gamma_{\mathrm{tx},j}^{*}A_{0}(\Delta_{ij})\right\} .
\end{align*}

The elements of the lower off diagonal are obtained as

\begin{align*}
\Phi(\theta_{\mathrm{tx},i},\tau_{j}) & =-\frac{1}{\sigma^{2}}\Re\left\{ h_{i}^{*}h_{j}\dot{\gamma}_{\mathrm{tx},i}^{*}\beta_{ij}\gamma_{\mathrm{tx},j}^{*}A_{1}(\Delta_{ji})\right\} \\
\Phi(\theta_{\mathrm{rx},i},\tau_{j}) & =-\frac{1}{\sigma^{2}}\Re\left\{ h_{i}^{*}h_{j}\gamma_{\mathrm{tx},i}\dot{\beta}_{ji}^{*}\gamma_{\mathrm{tx},j}^{*}A_{1}(\Delta_{ji})\right\} \\
\Phi(h_{R,i},\tau_{j}) & =-\frac{1}{\sigma^{2}}\Re\left\{ h_{j}\gamma_{\mathrm{tx},i}\beta_{ij}\gamma_{\mathrm{tx},j}^{*}A_{1}(\Delta_{ji})\right\} \\
\Phi(h_{I,i},\tau_{j}) & =-\frac{1}{\sigma^{2}}\Re\left\{ jh_{j}\gamma_{\mathrm{tx},i}\beta_{ij}\gamma_{\mathrm{tx},j}^{*}A_{1}(\Delta_{ji})\right\} \\
\Phi(\theta_{\mathrm{rx},i},\theta_{\mathrm{tx},j}) & =\frac{1}{\sigma^{2}}\Re\left\{ h_{i}^{*}h_{j}\gamma_{\mathrm{tx},i}\dot{\beta}_{ji}^{*}\dot{\gamma}_{\mathrm{tx},j}A_{0}(\Delta_{ij})\right\} \\
\Phi(h_{R,i},\theta_{\mathrm{tx},j}) & =\frac{1}{\sigma^{2}}\Re\left\{ h_{j}\gamma_{\mathrm{tx},i}\beta_{ij}\dot{\gamma}_{\mathrm{tx},j}A_{0}(\Delta_{ij})\right\} \\
\Phi(h_{I,i},\theta_{\mathrm{tx},j},) & =\frac{1}{\sigma^{2}}\Re\left\{ jh_{j}\gamma_{\mathrm{tx},i}\beta_{ij}\dot{\gamma}_{\mathrm{tx},j}A_{0}(\Delta_{ij})\right\} \\
\Phi(h_{R,i},\theta_{\mathrm{rx},j}) & =\frac{1}{\sigma^{2}}\Re\left\{ h_{j}\gamma_{\mathrm{tx},i}\dot{\beta}_{ij}\gamma_{\mathrm{tx},j}^{*}A_{0}(\Delta_{ij})\right\} \\
\Phi(h_{I,i},\theta_{\mathrm{rx},j}) & =\frac{1}{\sigma^{2}}\Re\left\{ jh_{j}\gamma_{\mathrm{tx},i}\dot{\beta}_{ij}\gamma_{\mathrm{tx},j}^{*}A_{0}(\Delta_{ij})\right\} \\
\Phi(h_{I,i},h_{R,j}) & =\frac{1}{\sigma^{2}}\Re\left\{ j\gamma_{\mathrm{tx},i}\beta_{ij}\gamma_{\mathrm{tx},j}^{*}A_{0}(\Delta_{ij})\right\} .
\end{align*}

\subsubsection*{Remarks}
\begin{itemize}
\item When $p(t)$ is flat in the frequency domain, then
\begin{align*}
A_{0}(\Delta) & =\frac{\sin(\pi B\Delta)}{\pi B\Delta}\\
A_{1}(\Delta) & =\frac{-\sin(\pi B\Delta)+\pi B\Delta\cos(\pi B\Delta)}{\pi B\Delta^{2}}.\\
A_{2}(\Delta) & =\frac{((\pi B\Delta)^{2}-2)\sin(\pi B\Delta)}{\pi\Delta^{3}B}\\
 & +\frac{2\pi B\Delta\cos(\pi B\Delta)}{\pi\Delta^{3}B}.
\end{align*}
We observe that $A_{0}(0)=1$, $A_{1}(0)=0$, and $A_{2}(0)=\pi^{2}B^{2}/3$,
so that the entries in $\mathbf{J}_{(i,i)}$ have compact expressions,
compared to $\mathbf{J}_{(i,j\neq i)}$. 
\item When $\Delta$ is such that $B\Delta\gg1$, then $A_{0,1,2}(\Delta)\approx0$.
Hence, when paths have large relative path lengths, this leads to
a block diagonal structure in (\ref{FIM_gral}). 
\end{itemize}

\section{Optimized Beam Directions\label{sec:Optimized-Beam-Directions}}

The directions of the beams in the position-aided protocol can be
optimized as follows. We select one beam with maximum gain direction
$\theta_{0}$ closest to $\hat{\theta}_{\mathrm{tx},0}^{(i)}$. Then,
the two additional beams are set to minimize a measure of expected
future uncertainty. Let $\mathbf{J}_{\hat{\bm{\eta}}}^{(i,m)}(\theta)$
be the FIM, evaluated in $\hat{\bm{\eta}}$ (the estimate of $\boldsymbol{\eta}$)
for a beam pointing towards $\theta$. The we choose beams 
\begin{align*}
\mathrm{minimize}_{\theta_{1},\theta_{2}}\, & \mathrm{trace}\Bigl\{\Bigl[\mathbf{J}_{\hat{\bm{\eta}}}^{(i,1)}(\theta_{0})\\
 & +\mathbf{J}_{\hat{\bm{\eta}}}^{(i,2)}(\theta_{1})+\mathbf{J}_{\hat{\bm{\eta}}}^{(i,3)}(\theta_{2})\Bigr]_{1:2,1:2}^{-1}\Bigr\}\\
\mathrm{subject\,to}\,\, & \theta_{1},\theta_{2}\in\Theta.
\end{align*}
In case $\Theta$ is $[-\pi/2,\pi/2]$, we can instead set $\theta_{0}=\hat{\theta}_{\mathrm{tx},0}^{(i)}$
, $\theta_{1}=\hat{\theta}_{\mathrm{tx},0}^{(i)}+\varepsilon$ and
$\theta_{2}=\hat{\theta}_{\mathrm{tx},0}^{(i)}-\varepsilon$, and
optimize with respect to the scalar parameter $\varepsilon\ge0$.

\bibliographystyle{IEEEtran}
\bibliography{references}

\end{document}